\def\C{{\bf C}}%
\def\P{{\bf P}}%
\def\Q{{\bf Q}}%
\def\Z{{\mathbb Z}}%
\def\negf{f_-}
\def\recipf{{f_{-1}}}

\def\psom#1{\mbox{\fbox{\ensuremath{#1}}}}

\documentclass{sig-alternate}
\usepackage[show]{ed}

\newtheorem{theorem}{Theorem}
\newtheorem{conjecture}{Conjecture}
\newtheorem{definition}{Definition}
\newtheorem{example}{Example}
\newtheorem{application}{Application}
\newtheorem{proposition}[theorem]{Proposition}
\newtheorem{notation}{Notation}
\newtheorem{corollary}[theorem]{Corollary}
\newtheorem{lemma}[theorem]{Lemma}
\newtheorem{claim}{Claim}
\usepackage{url}
\usepackage{listings}
\usepackage{comment}

\begin{document}
\title{The Power of Vocabulary:\\The Case of Cyclotomic Polynomials
\titlenote{Research begun while both authors were on sabbatical at David R.
Cheriton School of Computer Science,
University of Waterloo, Waterloo, Ontario}}
\numberofauthors{2}
\author{
\alignauthor
Jacques Carette 
\affaddr{Department of Computing and Software}\\
\affaddr{McMaster University,Hamilton, Ontario}\\
\email{carette@mcmaster.ca}
\alignauthor
James H.~Davenport
\affaddr{Department of Computer Science}\\
  \affaddr{University of Bath, Bath BA2 7AY, United Kingdom}\\
  \email{J.H.Davenport@bath.ac.uk}
}
\maketitle
\begin{abstract}
We observe that the vocabulary used to construct the ``answer'' to
problems in computer algebra can have a dramatic effect on the computational
complexity of solving that problem.  We recall a formalization of this
observation and explain the classic example of sparse polynomial arithmetic. 
For this case, we show that it is possible to extend the
vocabulary so as reap the benefits of conciseness whilst avoiding the obvious
pitfall of repeating the problem statement as the ``solution''.

It is possible to extend the vocabulary {\it either\/} by irreducible
cyclotomics {\it or\/} by $x^n-1$: we look at the options and suggest that the
pragmatist might opt for both.
\end{abstract}

%\section{Complexity depends on vocabulary}
%\section{Extending Vocabulary wins}
%Depends on Jacques' 2004 paper \cite{Carette2004}.
%\section{Even fixed differences matter}
%\subsection{Dense/Sparse for Polynomials/Matrices}
%\subsection{Further Polynomials --- SLP etc.}
%\subsection{Term Equivalence for Groups}
%\section{What about factoring of sparse polynomials?}

\section{Introduction}
While sparse polynomials are a natural data structure for human beings (who
writes $x^{10}+0x^9+0x^8+0x^7+0x^6+0x^5+0x^4+0x^3+0x^2+0x^1-1$?) and computer
algebra systems, algorithms to do more than add and multiply are scarce on the
ground, and most texts slip silently from considering sparse polynomials to
considering dense ones \cite{DavenportCarette2010}. This is partly because of
the existence of examples showing that the
output can be exponentially larger than the input, and hence ``nothing can be
done''. We contend that these examples are basically all cases of the
cyclotomic polynomials in disguise, and that, by admitting these to the
{\it output\/} language, as Schinzel's $K$-operator \cite{Schinzel1982}
effectively does, these examples cease to be absolute
barriers to efficient algorithms. Cyclotomic factors can often be recognised
relatively efficiently \cite{BradfordDavenport1989}, though the worst-case is
NP-hard from this result. 
\def\foo{\cite[Theorem 6.1]{Plaisted1977}}
\begin{theorem}[\foo]It is NP-hard to solve the problem, given a polynomial
$p(x)\in\Z[x]$, to determine if $p$ has a root $r$ of modulus
1.\label{thm:Plaisted3}
\end{theorem}

This is a paradigmatic example of a more general thesis: solving problems
in computer algebra requires the concurrent design of the most appropriate
vocabulary and algorithms which are polynomial {\it in the size of the
output\/} so encoded.  Naturally, unconstrained multiplication of new
vocabulary is not a viable solution, and a methodology for costing this was
proposed in \cite{Carette2004}.

A common problem in computer algebra is ``factorize this polynomial''. The
algorithms commonly used first compute factorizations $p$-adically, and then
deduce the ``true'' factorization over $\Z$. The traditional approaches
\cite{Zassenhaus1969} are theoretically exponential in the number of $p$-adic
factors, though in practice the exponential aspect can be ``controlled''
\cite{Abbottetal2000a}. Polynomial-time (in the degree, and {\it a fortiori\/}
in the number of $p$-adic factors) algorithms are known
\cite{Lenstraetal1982},  but in practice tend to be slower. The most recent
progress is in \cite{vanHoeij2002}, whose algorithm is faster in practice, and
the deduction phase is heuristically polynomial time in the number of $p$-adic
factors.

Of course, for a sparse polynomial such as $x^n-1$, the size of the polynomial
is $O(\log n)$, and so an algorithm polynomial in $n$ is still exponential in
the size of the input.  Are there algorithms which are polynomial in the size
of the input?  If the output represents the factors as expanded polynomials,
this is impossible.  There is however a conjecture that the only cases
which cause exponential blowups are cyclotomic factors -- we will return to
this later.

%\subsection{Complexity depends on vocabulary}
%
%Be more precise about this.
%
%\subsection{Contribution}
%
%So what is this paper about?
%
%\subsection{Outline}
%
%the plan.

\begin{notation}We define the following
for a polynomial
$$f=\sum_{i=0}^na_ix^i=a_n\prod_{i=1}^n(x-\alpha_i):$$
\end{notation}
% JHD: I find the repeated use of italic and mathematics confusing
\begin{description}
\item[$\#f $]the number of non-zero terms in $f$, $|\{i: a_i\ne0\}|$;
\item[$\recipf$]$=x^nf(1/x)=\sum_{i=0}^na_{n-i}x^i$;
\item[$\negf$]$=f(-x)=\sum_{i=0}^na_i(-x)^i$;
\item[$f_e$]the even part of $f$, $\sum_{i=0}^{n/2}a_{2i}x^i$;
\item[$f_o$]the odd  part of $f$, $\sum_{i=0}^{n/2}a_{2i+1}x^i$;
\item[$f_2$]the root-square, or Graeffe\footnote{There are various
conventions in the literature as to how one handles the $\pm$ arising from the
parity of $n$.}, of $f$, $\pm
a_n^2\prod_{i=1}^n(x-\alpha_i^2)= f_e^2-xf_o^2$;
\item[$||f||_\infty$]$=\max_{i=0}^n|a_i|$;
\item[$||f||_2$]$=\sqrt{\sum_{i=0}^n a_i^2}$;
\item[$||f||_1$]$={\sum_{i=0}^n |a_i|}$;
\item[$M(f)$](the Mahler measure of $f$) $=|a_n|\prod_{i=1}^n\max(1,|\alpha_i|)$.
\end{description}

\begin{table*}[t]
\caption{Large coefficients in $\Phi_k$}\label{tab:phi}
\begin{tabular}{lrrrrrrrrrrrrr}
$|a_i|$&2&3&4&5&6&7&8=9&14&23&25&27&59&359\\
first
$\Phi_k$&105&385&1365&1785&2805&3135&6545&10465&11305&17225&20615&26565
&40755\\
$\phi(k)$&48&240&576&768&1280&1440&3840&6336&6912&10752&12960& 10560&17280\\
\end{tabular}
% Complete up to 80,000: JHD on 14.4.2009
\hfil\break 
[A larger version, independently computed, is in \cite{ArnoldMonagan2008}.]
\end{table*}
\section{Cyclotomic Polynomials}
\begin{notation}
Let $d(n)$ denote the number of divisors of the number $n$ (including 1 and
$n$ itself).
\end{notation}
\begin{theorem}\label{thm:Wigert}
It is known \cite{Wigert1907} that
\begin{equation}
d(n)\le n^{(\log 2+o(1))/\log\log n}.
\end{equation}
\end{theorem}
However, we should note the caveats about the distribution of $d(n)$ given at
\cite[Theorem 432]{HardyWright1979}, in particular that it has {\it average\/}
order $\log n$ but {\it normal\/} order roughly $(\log n)^{\log 2}$.
\begin{definition}
We will say that a polynomial is {\em cyclotomic\/}, if all its roots are
roots of unity. \rm Many authors reserve this for irreducible polynomials, but
we will explicitly say ``irreducible'' when we need to.
\end{definition}
\begin{notation}
Let $\Phi_k$ be the $k$-th irreducible cyclotomic polynomial:
\begin{equation}
\Phi_k(x)=\prod_{\gcd(j,k)=1}(x-e^{2\pi ij/k}).
\end{equation}
We denote by $C_n$ the cyclotomic polynomial with all $n$-th roots of 
unity, i.e. $C_n = x^n-1$.
\end{notation}
We should note that it is {\it not\/} the case that the coefficients of
$\Phi_k$ are 0 or $\pm1$. The first counterexample is $\Phi_{105}$, which
contains the terms $-2x^7$ and $-2x^{41}$.
% $\Phi_{1155}$ 
%contains the terms $3x^{95}$, $3x^{115}$, $3x^{117}$, $3x^{146}$, $3x^{229}$,
%$3x^{240}$, $3x^{251}$, $3x^{334}$, $3x^{363}$, $3x^{365}$ and $3x^{385}$, as
%well as $-3x^{94}$, $-3x^{116}$, $-3x^{194}$, $-3x^{286}$, $-3x^{364}$ and
%$-3x^{386}$. 
$\Phi_{385}$ contains the terms $-3x^{120}$ and $-3x^{121}$ 
The growth rate is in fact greater than one might expect, and
$\Phi_{15015}$ has terms of $23\,{x}^{2294}$ and $23\,{x}^{3466}$.
% -532, +500 in 15015*17; -1182,1167 in 15015*19, -1311, 1218 in 15015*23
$15015=3\cdot5\cdot7\cdot11\cdot13$, and this looks like the recipe 
(confirmed in \cite{Batemanetal1984}) to make
large values\footnote{It {\it does\/} give us (exactly) 500 at
$255255=3\cdot5\cdot7\cdot11\cdot13\cdot17$.}, but in fact 23 is first
attained at $11305=5\cdot7\cdot17\cdot19$, as shown in table \ref{tab:phi}.
We note the spectacular leap at $40755=3\cdot5\cdot11\cdot13\cdot19$, which is
the largest coefficient up to $k=80,000$. \cite[Table 3]{ArnoldMonagan2008}
shows more such leaps for (much) larger $n$. 
\begin{theorem}\label{thm:vaughan}
\cite[Theorem 1]{Vaughan1974} shows that, for infinitely many $n$,
\begin{equation}\label{eq:Phi}
\log||\Phi_n||_\infty>\exp\left(\frac{(\log 2)(\log n)}{\log\log n}\right),
\end{equation}
\end{theorem}
and indeed this is precisely the right order of (worst-case) growth 
\cite{Bateman1949}, perhaps better expressed as
\begin{equation}\label{eq:Phigrowth}
\limsup_{n\rightarrow\infty}\frac{||\Phi_n||_\infty}{\log n/\log \log n}=\log 2.
\end{equation}
\begin{proposition}$x^n-1=\prod_{d|n}\Phi_d(x)$, and these factors are
irreducible.
\end{proposition}
\begin{proposition}\label{prop:Mobius}
$\Phi_n(x)=\prod_{d|n}(x^d-1)^{\mu(n/d)}$, where $\mu$ is the
M\"obius function.
\end{proposition}
\begin{proposition}$x^n-1$ has $d(n)$ irreducible factors.
\end{proposition}
Cyclotomic polynomials are the bugbear of anyone who tries to deal with sparse
polynomials. 
\begin{example}\label{ex:factor}
Asking for the factorization, or even the degrees of the factors, of $x^n-1$
is tantamount to factoring $n$, since for every prime $p$ dividing $n$, there
is an $x^{p-1}+\cdots$ in the factorization of $x^n-1$.
\end{example}
\begin{example}\label{ex:phifact}
Similarly, asking for the degree of $\Phi_k$ is, if $k$ is $p\cdot q$ ($p$,
$q$ distinct primes), tantamount to factoring $k$, since $\phi(k)=(p-1)(q-1)$
and so
$$
p,q = \frac12\left(k+1-\phi(k)\pm\sqrt{k^2-2k-2k\phi(k)+(\phi(k)-1)^2}\right).
$$
\end{example}
Cyclotomic polynomials are frequently used as examples.
\begin{example}\cite[p. 185]{vanHoeij2002} gives this example
$$
{x}^{128}-{x}^{112}+{x}^{80}-{x}^{64}+{x}^{48}-{x}^{16}+1,
$$
and states that his algorithm sped up Maple by a factor of 500 on this
example. From a cyclotomic-aware point of view, such as
\cite{BradfordDavenport1989}, this polynomial is easy. Four applications of
Graeffe's root-squaring process show (as is obvious to the eye) that this is
$f(x^{16})$ where
$$
f(x) = {x}^{8}-{x}^{7}+{x}^{5}-{x}^{4}+{x}^{3}-x+1.
$$
Another application takes $f$ to itself, and hence $f$, and so the original
polynomial, is cyclotomic. If $\alpha$ is a root of $f$, $\alpha^{15}=1$, so
$f=\Phi_{15}$, and the original polynomial is 
$$
\Phi_{15}\Phi_{30}\Phi_{60}\Phi_{120}\Phi_{240}.
$$
\end{example}
\begin{example}
If $p$ is prime, $f=x^p-1$, then $\#f=2$ but $f=(x-1)(x^{p-1}+\cdots+1)$: two
factors with 2 and $p$ terms respectively.
\end{example}
\begin{example}\label{ex:twofact}
If $p$, $q$ are distinct primes, $f=(x^p-1)(x^q-1)$, then $\#f=4$ but
$f=(x-1)^2(x^{p-1}+\cdots+1)(x^{q-1}+\cdots+1)$. The square-free
decomposition of $f$ is therefore a repeated factor with 2 terms and a factor
$g$ with $pq-p-q+2$ terms respectively. The largest coefficient in $g$ is
$\min\{p,q\}$, and $||g||_1=pq$.
\end{example}
Obviously, a square-free decomposition of
$f$ was a bad idea in this case: however previously-proposed algorithms, e.g.
\cite[p. 69]{BeukersSmyth2002} tend to do this.
\par
It could be argued that the problem in this case is the `cofactor', but life
is not that simple.
\begin{example}
If $p$, $q$ are distinct primes, $f=(x^p-1)^2(x^q-1)$, then $\#f=6$ but the
square-free factorization is
$$
(x-1)^3\left(x^{p-1}+\cdots+1\right)^2\left(x^{q-1}+\cdots+1\right),
$$
and we are forced to write out the large squared factor. The largest
coefficient of $\left(x^{p-1}+\cdots+1\right)^2$ is $p$, so we had also better
not compute it and then take its square root.
\end{example}
It is the contention of this paper that all these difficulties {\it except the
first\/} are caused by an inadequate vocabulary: the first seems to be
intrinsic, in the fact the factorization of numbers can be encoded as a
problem of factorization of polynomials. All we can do is recognise the fact.
\section{Representational Complexity}\label{sec:RC}
Information theory, whether through the guise of Kolmogorov Complexity
\cite{LiVitanyi} or Minimum Description Length \cite{MDLbook}, tell us that
\emph{good} representations of structured objects are two-part codes: a model
and an encoding of data using that model.  In other words, the proper
``length'' of an object consists in counting the length of the representation
of the model as well as the representation of the data encoded using this
model.  
\par
In \cite{Carette2004}, these results from information theory are rephrased
so as to apply more directly to Computer Algebra Systems, and applications
to simplification are outlined.  The basic result is that for large enough
structured expressions, it is \emph{always} worthwhile to first formalize
the ``structure'', and then encode the data in such a way as to abstract out
that structure.  Note that, if model extensions are not allowed, then 
simplification reduces simply to length reduction.  It is exactly the 
confusion between issues of the (background) model-class and its use in
model reduction which caused Moses \cite{Moses1971a} to argue
that ``simplification'' was impossible to formalize.
\par
When tackling a particular situation, \cite{Carette2004} boils down to
finding the right \emph{vocabulary} in which to express ones' result.  In
some cases, the right vocabulary is somewhat counter-intuitive.  For example,
in the case of algebraic numbers, it was long ago discovered that using
minimal polynomials to ``encode'' an algebraic number was best -- although
this can seem puzzling in the setting of ``solving'' a polynomial, as then
the answer to the problem is just an encoding of the question.  We will
return to this issue later.
\par
For the particular case of factoring of polynomials, what does this tell us?
All the theoretical results point in the same direction: cyclotomic polynomials
are somehow ``special cases'', especially when one is factoring sparse
polynomials.  Conventional wisdom already tells us that both dense and
sparse polynomials are useful model classes, and that we should have both
at our disposal.  The mathematical theory of factoring polynomials (as outlined
in the rest of this paper) informs us that cyclotomics are undeniably part
of the domain of discourse.  Combining these together tells us that they 
should also be part of our model classes.  The only remaining question 
then is whether adding this particular vocabulary actually leads to
a \emph{simplification}.  To evaluate this, we need to actually display
some data structures designed with this new vocabulary, and then evaluate
if we have made any real gains.
\section{Data Structures}\label{sec:DS}
Since we will be arguing on the size of data structures, we will need to define
our data representations. The precise details might vary, though in practice
the conclusions will not. For definiteness, we describe our choices according
to the {\it unaligned\/} packed encoding of ASN.1 \cite{ITU2002}: note that
their {\tt SEQUENCE} is what C programmers would think of as {\tt struct}.
Our encodings are intended to be practical, though we ignore issues of
alignment to word boundaries, and indeed a number of $\lceil\ldots\rceil$
operators are also omitted.
\par
In theory, one needs to have arbitrary sized data fields, which means one needs
fields for the length of the size fields. To avoid this, we will
assume that $N$ and $K$ are global parameters for the size of the object,
bounding the degrees and the size of the coefficients. Since a polynomial's
factors always have smaller degree, we can assume $N=\log_2n$. It is not so
easy for the coefficients \cite{Mignotte1981a}, but we will assume a {\it
single\/} field of $K$ bits associated with each outermost data structure,
giving the size $k$ of all the coefficients stored in that structure. Since
there is one of these, we can ignore its cost.
\par
We next give explicit representations for dense polynomials, sparse
polynomials, factored polynomials, $\Phi$-aware factorizations and 
$C$-aware factorizations (explained fully below).

\subsection{A single dense polynomial}
We choose a dense representation with a uniform size bound for all the
coefficients.
A single dense polynomial of degree $n$ requires $\log_2 n$ bits to represent
the degree, and then there are $n+1$ coefficients. Hence if each of them
requires $k$ bits, we need $\log_2 k$ bits for the telling us this, and then
the coefficients require $k+1$ bits (including sign).
\begin{equation}
(k+1)(n+1)+\log_2 k+\log_2 n
\label{Dense1}
\end{equation}
In pseudocode\footnote{Essentially ASN.1, except that we allow ourselves to write mathematics, enclosed in boxes, in the pseudocode.}, this might be represented as follows.
\begin{lstlisting}[mathescape]
DensePoly ::= SEQUENCE {
    Degree INTEGER ($\psom{0\ldots 2^N-1}$),
    k      INTEGER ($\psom{0\ldots 2^K-1}$),
    Coefficients SEQUENCE {
           INTEGER  ($\psom{-2^k\ldots 2^k-1}$)
           } SIZE ($\psom{\hbox{\tt Degree}+1}$) 
    }
\end{lstlisting}
\subsection{A single sparse polynomial}
We choose a sparse representation with a uniform size bound for all the
coefficients. Furthermore, we assume that there are $t$ non-zero coefficients,
i.e. $t$ terms to be represented, and $t$ is bounded by the same bound as the
degree.
A single sparse term from a polynomial of degree $n$ requires $\log_2 n$ bits to represent the degree. Hence the total space is given by
\begin{equation}
%\log_2 n + t(\log_2 k+ 1+\log_2 n).
% JHD 22/1/2010: BUG should be k not \log_2k
\log_2 n + t(k+ 1+\log_2 n).
\label{Sparse1}
\end{equation}
In pseudocode, this might be represented as
\begin{lstlisting}[mathescape]
SparsePoly ::= SEQUENCE {
    TermCount INTEGER ($\psom{0\ldots 2^N-1}$),
    Terms SEQUENCE {
        Degree INTEGER ($\psom{0\ldots 2^N-1}$),
        Coefficient INTEGER  ($\psom{-2^k\ldots 2^k-1}$)
        } SIZE (TermCount)
    }
\end{lstlisting}
Using a Horner scheme might save a few more bits in the representation of
the exponent, but rarely appreciably so.
\subsection{Representing Factorizations}\label{sec:mult}
We will use the same structure for square-free or complete factorizations: the
number of (distinct) factors followed by pairs (multiplicity, factor).
In pseudocode, this might be represented as
\begin{lstlisting}[mathescape]
Factorization ::= SEQUENCE {
    FactorCount INTEGER ($\psom{0\ldots 2^N-1}$),
    Factors SEQUENCE {
        Multiplicity INTEGER ($\psom{0\ldots 2^N-1}$),
        Factor $\psom{Poly}$
        } SIZE (FactorCount)
    }
\end{lstlisting}
where $Poly$ is one of \verb+DensePoly+ or \verb+SparsePoly+.
Hence with $f$ factors, the overhead (i.e. the cost over and above that of
storing the distinct factors themselves) is
\begin{equation}
(f+1)\log_2 n.
\end{equation}
\subsection{Representing $\Phi$-aware Factorizations}\label{sec:pmult}
We will use the same structure for square-free or complete factorizations: the
number of (distinct) factors followed by pairs (multiplicity, factor).
However, we do this twice: once for the factors that are $\Phi_k$, and once for
those that are not. The factors that {\it are\/} $\Phi_k$ are stored as
$k$ followed by\footnote{The inclusion of $\phi(k)$ avoids the problem in
example \ref{ex:phifact}.}  $\phi(k)$.
In pseudocode, this might be represented as
\begin{lstlisting}[mathescape]
Factorization ::= SEQUENCE {
    PhiFactorCount INTEGER ($\psom{0\ldots 2^N-1}$),
    PhiFactors SEQUENCE {
        Multiplicity INTEGER ($\psom{0\ldots 2^N-1}$),
        k INTEGER ($\psom{0\ldots 2^N-1}$),
        Degree INTEGER ($\psom{0\ldots 2^N-1}$),
        } SIZE (PhiFactorCount),
    FactorCount INTEGER ($\psom{0\ldots 2^N-1}$),
    Factors SEQUENCE {
        Multiplicity INTEGER ($\psom{0\ldots 2^N-1}$),
        Factor $\psom{Poly}$
        } SIZE (FactorCount)
    }
\end{lstlisting}
where $Poly$ is one of \verb+DensePoly+ or \verb+SparsePoly+.
Hence with $l$ $\Phi_k$ factors, the cost of storing them is $(3l+1)\log n$.
% JHD Although I wrote this (I think) I don't understand it - below is what I might have meant.
%Of course, a polynomial stored this way is cheaper than {\it any\/} polynomial
%stored in the previous representation, so the total cost is bound to be less.
% JC: the text below makes sense, while that above is hard to fathom.
{\it Any\/} $\Phi_k$ stored this way is cheaper than in any of the previous
representation (dense, sparse or factored), so the worst case is when there are
no $\Phi_k$ in the factorization, when the overhead is merely the one field
\verb+PhiFactorCount+.
\subsection{Representing $C$-aware Factorizations}\label{sec:cmult}
Instead of storing the $\Phi_k$, we could store the complete cyclotomic
polynomials $x^k-1$.
Again, we will use the same structure for square-free or complete
factorizations: the
number of (distinct) factors followed by pairs (multiplicity, factor).
Also in this case, we do this twice:
once for the factors that are of the form $x^k-1$, and once for
those that are not. One might think to store the factors that {\it are\/}
$x^k-1$ simply as $k$. However, as pointed out in  example \ref{ex:factor},
this will not allow us to answer questions such as ``how many factors'' in a
reasonable time. Hence we store $k$ followed by its prime factorization.
\par
We should also note that Proposition \ref{prop:Mobius} means that
we now need {\it negative\/} multiplicities as well.
In pseudocode, this might be represented as
\begin{lstlisting}[mathescape]
Factorization ::= SEQUENCE {
    CFactorCount INTEGER ($\psom{0\ldots 2^N-1}$),
    CFactors SEQUENCE {
        Multiplicity INTEGER ($\psom{-2^N\ldots 2^N-1}$),
        Degree INTEGER ($\psom{0\ldots 2^N-1}$),
        NumFactors INTEGER ($\psom{0\ldots 2^N-1}$),
        KFactor SEQUENCE {
            INTEGER ($\psom{0\ldots 2^N-1}$)
            } SIZE (NumFactors),
        } SIZE (CFactorCount),
    FactorCount INTEGER ($\psom{0\ldots 2^N-1}$),
    Factors SEQUENCE {
        Multiplicity INTEGER ($\psom{0\ldots 2^N-1}$),
        Factor $\psom{Poly}$
        } SIZE (FactorCount)
    }
\end{lstlisting}
where $Poly$ is one of \verb+DensePoly+ or \verb+SparsePoly+.
Hence with $f$ factors of $x^k-1$ involved, the overhead (i.e. the cost over
and above that of storing the distinct factors themselves) is
\begin{equation}
(f+1)\log_2 n.
\end{equation}
\par
An alternative formulation might store the factors of $k$ with multiplicity:
there does not seem to be a great deal to choose between them. In either case,
we should note that asking for the {\it number\/} of irreducible factors is no
longer trivial, since the number of irreducible factors corresponding to a
single $x^k-1$ is $d(k)$. Furthermore, since we are allowed negative
exponents, representing 
$$x^k+1 :=\left(x^{2k}-1\right)/\left(x^k-1\right),$$
we have to note that not all the irreducible factors of $x^{2k}-1$ are
actually factors of the left-hand side. Nevertheless, since we have stored the
prime factorization of $k$, the problem is efficiently soluble (certainly
polynomial time in the size of the representation).
\begin{table*}[th]
\caption{\label{table:cyc}Representing $x^n-1=\prod_{d|n}\Phi_d(x)$}
\begin{tabular}{rrrr}
Representation&Fully expanded&Square-free factorization&Factored\\
Dense&$2(n+1)+\log_2 n$&same as factored&
%$\left(n+2d(n)\right)\log_2n+\cdots$
$n^{1+\frac{\log 2}{\log\log n}}\log_2e$
\\
Sparse&$3\log_2 n$&same as factored&
%$\left(2n+3d(n)\right)\log_2n+\cdots$
$n^{1+\frac{\log 2}{\log\log n}}\log_2e$
\\
$\Phi_k$&$3\log_2 n$&same as factored&$(2d(n)+1)\log_2 n$
\\
$x^k-1$&$3\log_2 n$&same as factored&$\log_2 n$
\\
%\\
\end{tabular}
\end{table*}
\section{Representing some cyclotomic polynomials}\label{sec:cycrep}
\begin{table*}[t]
\caption{Representing $(x^p-1)(x^q-1)=(x-1)^2\Phi_p(x)\Phi_q(x)$
with $n=p+q$\label{table:cyc2}}
\begin{tabular}{rrrr}
Representation&Fully expanded&Square-free factorization&Factored\\
Dense&$2(n+1)+\log_2 n$&$(1+\log_2 n)(n+2)+4\log_2 n$&$2(n+3)+6\log_2 n$
\\
Sparse&$4\log_2 n$&$(2n+2)\log_2 n$&$(n+10)\log_2n$
\\
$\Phi_k$&$4\log_2 n$&$6\log_2 n$&$6\log_2 n$
\\
$x^k-1$&$4\log_2 n$&$2\log_2 n$&$2\log_2 n$
\\
\end{tabular}
\end{table*}
We now use each of our representations and compute the size of the results.
\subsection{Factorization of $x^n-1$}\label{sec:dense1}
The sizes of the expanded $x^n-1$ polynomial in dense, and sparse encodings
are obvious.  The $\Phi$-aware and $C$-aware versions are within an
additive constant of \verb+DensePoly+ / \verb+SparsePoly+.  For definiteness,
we will use \verb+SparsePoly+ based counts.
\par
To understand the size of the factored forms, we need to study their
sizes a little more closely.  In this case, the factored form and the 
square-free form coincide.
There are $d(n)$ factors, of total degree $n$, hence $n+d(n)$ terms.
By Theorem \ref{thm:vaughan}, we can bound\footnote{In theory, not all the
factors {\it can\/} have coefficients this large, but the gain from exploiting
this is relatively small.} the size of the coefficients as
$\log_2e$ times the right-hand side of (\ref{eq:Phi}), {\it viz.\/} $\frac{\log
n}{\log\log n}$, to get the size of a dense polynomial to be
%\begin{equation}
%(n+d(n))\left(1+\log n\left(\log_2e+\frac1{\log\log n}\right)\right)+
%d(n)\log_2n.
%\end{equation}
$$
%(n+d(n))\left(1+\log_2 n\left(1+\frac{\log 2}{\log\log n}\right)\right)+
%d(n)\log_2n
%=\left(n+2d(n)\right)\log_2n+\cdots .
(n+d(n))\log_2e\exp\left(\frac{(\log 2)(\log n)}{\log\log n}\right)+
d(n)\log_2n
$$
\begin{equation}\label{eq:densephi}
\approx n^{1+\frac{\log 2}{\log\log n}}\log_2e.
\end{equation}
\par
{\it In general\/}\footnote{$x^{2^k}-1$ is an obvious counter-example.}, the
factors will be essentially dense, so a sparse encoding will save nothing, but
have to pay for the cost of storing the degrees with each coefficient, adding
$(n+d(n))\log_2n$, to give
$$
%(n+d(n))\left(2+\log_2 n\left(1+\frac{\log 2}{\log\log n}\right)\right)+
(n+d(n))\left(\log_2n+\log_2e\exp\left(\frac{(\log 2)(\log n)}{\log\log n}\right)\right)+
d(n)\log_2n
$$
\begin{equation}\label{eq:sparsephi}
%=\left(2n+3d(n)\right)\log_2n+\cdots .
\approx n^{1+\frac{\log 2}{\log\log n}}\log_2e.
\end{equation}
We should note that the asymptotically dominant term is the coefficient
storage in this model, which is contrary to intuition, and even the experimental
data in table \ref{tab:phi}, but this merely shows that the asymptotics will
take time to be visible.
\par
The results of this section are summarised in Table \ref{table:cyc}.
%\subsection{Multiplicities}
%When we are handling square-free or complete factorisations of polynomials of
%degree $n$, we attach a multiplicity $\le n$ to each factor, costing us $\log
%n$ bits.
\subsection{A square-free factorization}\label{sec:sqfr1}
Let use now consider $(x^p-1)(x^q-1)$, with $p,q$ distinct primes.
The ``fully expanded'' versions are again obvious.
The square-free factorization of $(x^p-1)(x^q-1)=(x-1)^2\Phi_p(x)\Phi_q(x)$
involves multiplying out $\Phi_p(x)\Phi_q(x)$. This gives
us coefficients of size $O(n)$, in fact $n/2$ assuming that $p$, $q$ are
balanced, taking $(\log_2 n)-1$ bits to represent the magnitude\footnote{In
this case, they are all positive, but we can't count on this in general.}.
\par
In the factored representation, we have three factors, of degrees 1, $p-1$ and
$q-1$, i.e. total degree $n-1$. All coefficients are bounded by 1. Hence the
total is
\begin{equation}\label{eq:sparsesqfr}
(n-1+3)(\log_2n+2) + (3\cdot2+2)\log_2n \approx (n+10)\log_2n.
\end{equation}
\par
The results of this section are summarised in Table \ref{table:cyc2}.
\section{Implementation notes}
\subsection{Cyclotomic-free}
It is important to review the encodings of the previous section and notice
that for cyclotomic-free cases, these encodings involve constant overhead,
independent of the degree and of the number of factors.  In fact, by using
a bit or two in a header word (which modern computer algebra systems always
seem to use in their internal representations), one can choose between these
encodings as necessary.  In other words, cyclotomic-free polynomials do not
have to bear any extra representation cost for this vocabulary extension.
\par
We can also construct various mixed cases, in other words sparse polynomials
which factor into a cyclotomic part and a small dense cofactor.  The
difference in encoding cost is correspondingly mixed, although the
end result is similar: adding cyclotomics asymptotically wins.
\subsection{Which to choose?}
We have posited two encodings for ``cyclotomic-aware'' representations of
factorizations: one in terms of the irreducible cyclotomics $\Phi_k$ (section
\ref{sec:pmult}) and one in terms of the `complete' cyclotomics $C_k=x^k-1$
(section \ref{sec:cmult}).  Tables~\ref{table:cyc} and~\ref{table:cyc2} make it
clear that adding cyclotomics to ones' vocabulary is certainly
reprensentionally efficient.  
But which one should be used? We first summarize some
the advantages and disadvantages.
\begin{description}
\item[Pro $\Phi_k$: clearly polynomial]In the $\Phi_k$-representation,
the factorization of
$x^{p+q}+x^p+x^q+1$ is $\Phi_2^2\Phi_{2p}\Phi_{2q}$, where as in the
$C_k$-representation it is $C_{2p}C_p^{-1}C_{2q}C_q^{-1}$.  Functions
meant to extract information from products of polynomials (degrees,
multiplicity, etc) still function `easily', whereas spotting that the
$C_k$-representation is not square-free is cheap, but not `obvious'.
\item[Anti $\Phi_k$: worst-case blowup]The last two entries of table
\ref{table:cyc} show that, for highly composite $n$, representing $C_n$ in
terms of $\Phi_k$ can require almost $n$ times as much, i.e. exponentially
more, since the input has size $O(\log n)$, space, by Theorem
\ref{thm:Wigert}.
\item[Notes]The pragmatist would probably choose $\Phi$. The theoretician
would be swayed by the complexity argument and want $C$. Possibly the best
answer is to admit $\Phi$, and $C$ with an additional extension to the
vocabulary as in (\ref{eq:Cphi}).
\end{description}

To further illustrate one particular difficulty, let us consider factoring
of $C_{105}$.  It factors into
$$
{\Phi_{1}(x)}{\Phi_{3}(x)}
{\Phi_{5}(x)} {\Phi_{7}(x)}
{\Phi_{15}(x)} {\Phi_{21}(x)}
{\Phi_{35}(x)} {\Phi_{105}(x)}.
$$
But if all we have at our disposal is $C$, then the best we can do (which
is still better than using sparse polynomials) is
$$
C_1(x)\frac{C_3(x)}{C_1(x)}\frac{C_5(x)}{C_1(x)}\frac{C_7(x)}{C_1(x)}
\frac{C_{15}(x)C_1(x)}{C_3(x)C_5(x)}
\frac{C_{21}(x)C_1(x)}{C_3(x)C_7(x)}\times
$$
$$
\frac{C_{35}(x)C_1(x)}{C_5(x)C_7(x)}
\frac{C_{105}(x)C_{7}(x)C_{5}(x)C_3(x)}{C_{35}(x)C_{21}(x)C_{15}(x)C_1(x)}.
$$
An even more succinct representation is
\begin{equation}\label{eq:Cphi}
\prod_{\displaystyle k=1\atop k|105}^{105} \Phi_k(x)
\end{equation}
at the cost of another vocabulary extension.  This is why in the
representation of $C$ in section~\ref{sec:cmult}, we list the factors of
$n$, which allows us to recover this factorization relatively easily.
This still means that in the $\Phi$-aware encoding, obtaining
the number of factors, the degrees of each of the factors, or the
multiplicity of each of the factors is straightforward, while for the
$C$-aware encoding, these simple questions now require some (small) amount
of computation.
, and so forth.  All of these 
questions remain just as easy to answer in the $\Phi$-aware encodings as
they were before.  However, for the $C$-aware encodings, these ``simple''
questions now require some actual computations to resolve.  In other words, 
adding $\Phi$ to our vocabulary is a very minor change with clear
efficiency gains, while adding $C$ is slightly disruptive but with even
greater asymptotic efficiency gains.
\section{Conclusion}
In section~\ref{sec:cycrep} we have shown how some ``troublesome''
factorizations, i.e. examples \ref{ex:factor} and \ref{ex:twofact},
cease to consume inordinate space when the cyclotomics are represented
explicitly. But what of arbitrary polynomials and their factorizations?
\par
The answer is that we do not know, but there are some tantalizing results.
\cite{Filasetaetal2008} state that, provided $f$ is non-reciprocal ($f\ne\pm
f_{-1}$), $f$ has a factor with at most $c_2(\#f,||f||_\infty)$ terms,
independent of $n$, and quotes \cite{Schinzel1969} to say that if the
polynomial has no reciprocal factors, then all irreducible factors have at
most $c_2(\#f,||f||_\infty)$ terms. This is a significant step towards
controlling the dependence of the {\it output size\/} on $n$, though it
falls short of saying that  is polynomial in the following ways:
\begin{itemize}
\item there is no guarantee that $c_2$ depends {\it polynomially\/} on $\#f$
or $\log||f||_\infty$, and indeed $c_2$ is still rather mysterious to the
authors \cite[Challenge 4]{DavenportCarette2010};
\item nothing is said about the size of the coefficients, and all known
technology makes them depend exponentially on $n$ (i.e. the output size
depends linearly on $n$).
\end{itemize}
We note, though, that the {\it known\/} examples of such growth
\cite[$\S3$]{Mignotte1981a}, depend on cyclotomic polynomials, so one
could hope that the second problem does not occur in practice.
\par
We are convinced, and we hope to have convinced the reader, that, regardless
of whether using cyclotomics ultimately makes the problem of sparse polynomial
factorization more tractable, even if not guaranteed polynomial (see Theorem
\ref{thm:Plaisted3}), in the input size, that they are most definitely worth
having in the basic vocabulary used for the {\it output\/} of factoring.  We
strongly recommend that the specification for what it means to factor a
polynomial be thus amended.
\bibliography{../../../../../../jhd}
\end{document}